\documentclass[11pt]{book}

\usepackage{amssymb}
\usepackage{amsmath}
\usepackage{amsfonts}
\usepackage{epsfig}

\makeatletter

\renewenvironment{thebibliography}[1]
     {\section*{\bibname}%
      \@mkboth{\MakeUppercase\bibname}{\MakeUppercase\bibname}%
      \list{\@biblabel{\@arabic\c@enumiv}}%
           {\settowidth\labelwidth{\@biblabel{#1}}%
            \leftmargin\labelwidth
            \advance\leftmargin\labelsep
            \@openbib@code
            \usecounter{enumiv}%
            \let\p@enumiv\@empty
            \renewcommand\theenumiv{\@arabic\c@enumiv}}%
      \sloppy
      \clubpenalty4000
      \@clubpenalty \clubpenalty
      \widowpenalty4000%
      \sfcode`\.\@m}
     {\def\@noitemerr
       {\@latex@warning{Empty `thebibliography' environment}}%
      \endlist}

\makeatother

\newcommand{\sect}[1]{\setcounter{equation}{0}\section{#1}}

\renewcommand\bibname{References}

\begin{document}

\setlength{\baselineskip}{5.0mm}


\chapter[Supersymmetry -- Guhr]{Supersymmetry in Random Matrix Theory}
\thispagestyle{empty}

\ \\

\noindent
{{\sc Thomas Guhr}$^1$
\\~\\$^1$Fakult\"at f\"ur Physik, 
Universit\"at Duisburg--Essen,\newline 
Lotharstrasse 1, 47057 Duisburg, Germany}

\begin{center}
{\bf Abstract}
\end{center}
Supersymmetry is nowadays indispensable for many problems in Random
Matrix Theory. It is presented here with an emphasis on conceptual and
structural issues. An introduction to supermathematics is given. The
Hubbard--Stratonovich transformation as well as its generalization and
superbosonization are explained. The supersymmetric non--linear
$\sigma$ model, Brownian motion in superspace and the color--flavor
transformation are discussed.

\sect{Generating Functions}
\label{sec1}

We consider $N\times N$ matrices $H$ in the three symmetry
classes~\cite{Dys62} real symmetric, Hermitean or quaternion real,
that is, self--dual Hermitean. The Dyson index $\beta$ takes the
values $\beta=1,2,4$, respectively. For $\beta=4$, the $N \times N$
matrix $H$ has $2 \times 2$ quaternion entries and all its eigenvalues
are doubly degenerate.  For a given symmetry, an ensemble of random
matrices is specified by choosing a probability density function
$P(H)$ of the matrix $H$. The ensemble is referred to as invariant or
rotation invariant if
\begin{equation}
P(V^{-1}HV) = P(H)
\label{eq1.2}
\end{equation}
where $V$ is a fixed element in the group diagonalizing $H$, that is, in
${\rm SO}(N)$,  ${\rm SU}(N)$ or ${\rm USp}(2N)$ for $\beta=1,2,4$,
respectively. Equation~\eqref{eq1.2} implies that the probability
density function only depends on the eigenvalues,
\begin{equation}
P(H) = P(X) = P(x_1,\ldots,x_N) \ .
\label{eq1.3}
\end{equation}
Here, we write the diagonalization of the random matrix as $H=U^{-1} X
U$ with $X={\rm diag\,}(x_1,\ldots,x_N)$ for $\beta=1,2$ and $X={\rm
  diag\,}(x_1,x_1,\ldots,x_N,x_N)$ for $\beta=4$.  The $k$--point
correlation function $R_k(x_1,\ldots,x_k)$ measures the probability
density of finding a level around each of the positions
$x_1,\ldots,x_k$, the remaining levels not being observed. One
has~\cite{Dys62,Meh04}
\begin{equation}
R_k(x_1,\ldots,x_k) = \frac{N!}{(N-k)!}
       \intop_{-\infty}^{+\infty} dx_{k+1} \cdots
       \intop_{-\infty}^{+\infty} dx_N \, |\Delta_N(X)|^\beta P(X) \ ,
\label{eq1.6}
\end{equation}
where $\Delta_N(X)$ is the Vandermonde determinant.  If the
probability density function factorizes,
\begin{equation}
P(X) = \prod_{n=1}^N P^{(E)}(x_n) \ ,
\label{eq1.7}
\end{equation}
with a probability density function $P^{(E)}(x_n)$ for each of the
eigenvalues, the correlation functions~\eqref{eq1.6} can be evaluated
with the Mehta--Mahoux theorem~\cite{Meh04}. They are $k \times k$
determinants for $\beta=2$ and $2k \times 2k$ quaternion determinants
for $\beta=1,4$ whose entries, the kernels, depend on only two of
the eigenvalues $x_1,\ldots,x_k$.

Formula~\eqref{eq1.6} cannot serve as the starting point for the
Supersymmetry met\-hod.  A reformulation employing determinants is
called for, because these can be expressed as Gaussian integrals over
commuting or anticommuting variables, respectively.  The key object is
the resolvent, that is, the matrix $(x_p^--H)^{-1}$ where the
argument is given a small imaginary increment, $x_p^- = x_p - i
\varepsilon$. The $k$--point correlation functions are then defined as
the ensemble averaged imaginary parts of the traces of the resolvents
at arguments $x_1,\ldots,x_k$,
\begin{equation}
R_k(x_1,\ldots,x_k) = \frac{1}{\pi^k} \int P(H)
       \prod_{p=1}^k {\rm Im\,}{\rm tr\,}\frac{1}{x_p^--H} d[H] \ .
\label{eq1.8}
\end{equation}
The necessary limit $\varepsilon\to 0$ is suppressed throughout in our
notation. We write $d[ \cdot ]$ for the volume element of the quantity
in square brackets, that is, for the product of the differentials of
all independent variables.  The definitions~\eqref{eq1.6}
and~\eqref{eq1.8} are equivalent, but not fully identical.
Formula~\eqref{eq1.8} yields a sum of terms, only one coincides with
the definition~\eqref{eq1.6}, all others contain at least one $\delta$
function of the form $\delta(x_p-x_q)$, see Ref.~\cite{Guh98}.

Better suited for the Supersymmetry method than the correlation
functions~\eqref{eq1.8} are the correlation functions
\begin{equation}
\widehat{R}_k(x_1,\ldots,x_k) = \frac{1}{\pi^k} \int d[H] \, P(H)
                          \prod_{p=1}^k {\rm tr\,} \frac{1}{x_p-iL_p\varepsilon-H} 
\label{eq1.10}
\end{equation}
which also contain the real parts of the resolvents. The correlation
functions~\eqref{eq1.8} can always be reconstructed, but the way how
this is conveniently done differs for different variants of the
Supersymmetry method.  In Eq.~\eqref{eq1.8}, all imaginary increments
are on the same side of the real axis. In Eq.~\eqref{eq1.10}, however,
we introduced quantities $L_p$ which determine the side of the real
axis where the imaginary increment is placed. They are either $+1$ or
$-1$ and define a metric $L$. Hence, depending on $L$, there is an
overall sign in Eq.~\eqref{eq1.10} which we suppress.  We use the short
hand notations $x_p^\pm = x_p - iL_p\varepsilon$ in the sequel. In
some variants of the Supersymmetry method, it is not important where
the imaginary increments are, in the supersymmetric non--linear
$\sigma$ model, however, it is of crucial importance.  We return to
this point.

To prepare the application of Supersymmetry, one expresses the
correlation functions~\eqref{eq1.10} as derivatives
\begin{equation}
\widehat{R}_k(x_1,\ldots,x_k) =
              \frac{1}{(2\pi)^k} \frac{\partial^k}{\prod_{p=1}^k\partial J_p}
                                      Z_k(x+J) \Bigg|_{J_p=0}
\label{eq1.11}
\end{equation}
of the generating function
\begin{equation}
Z_k(x+J) = \int d[H] \, P(H) \prod_{p=1}^k 
                        \left(\frac{\det(H - x_p+iL_p\varepsilon - J_p)}
                                         {\det(H - x_p+iL_p\varepsilon + J_p)}\right)^\gamma
\label{eq1.12}
\end{equation}
with respect to source variables $J_p , \ p=1,\ldots,k$. For
$\beta=1,2$ one has $\gamma=1$ whereas $\gamma=2$ for $\beta=4$. For
later purposes, we introduce the $2k \times 2k$ matrices $x={\rm
  diag\,}(x_1,x_1,\ldots,x_k,x_k)$ and $J={\rm
  diag\,}(-J_1,+J_1,\ldots,-J_k,+J_k)$ for $\beta=2$ as well as the
$4k \times 4k$ matrices $x={\rm
  diag\,}(x_1,x_1,x_1,x_1\ldots,x_k,x_k,x_k,x_k)$ and $J={\rm
  diag\,}(-J_1,-J_1,+J_1,+J_1,\ldots,-J_k,-J_k,+J_k,+J_k)$ for
$\beta=1,4$, which appear in the argument of $Z_k$.  We write $x^\pm =
x-iL\varepsilon$.  Importantly, the generating function is normalized
at $J=0$, that is, $Z_k(x)=1$.

\sect{Supermathematics}
\label{sec2}

Martin~\cite{Mar59} seems to have written the first paper on
anticommuting variables in 1959. Two years later, Berezin introduced
integrals over anticommuting variables when studying second
quantization. His posthumously published book~\cite{Ber87} is still
the standard reference on supermathematics.

\subsection{Anticommuting Variables}
\label{sec2.1}

We introduce Grassmann or anticommuting variables $\zeta_p, \
p=1,\ldots,k$ by requiring the relation
\begin{equation}
\zeta_p \zeta_q = - \zeta_q \zeta_p \ , \qquad p,q=1,\ldots,k \ .
\label{eq2.1}
\end{equation}
In particular, this implies $\zeta_p^2 = 0$. These variables are
purely formal objects. In contrast to commuting variables, they do not
have a representation as numbers. The inverse of an anticommuting
variable cannot be introduced in a meaningful way. Commuting and
anticommuting variables commute. The product of an even number of 
anticommuting variables is commuting, 
\begin{equation}
(\zeta_p \zeta_q)\zeta_r =  \zeta_p \zeta_q \zeta_r = - \zeta_p \zeta_r \zeta_q
                                            = + \zeta_r \zeta_p \zeta_q = \zeta_r (\zeta_p \zeta_q) \ .
\label{eq2.1a}
\end{equation}
We view the anticommuting variables as complex and define a complex
conjugation, $\zeta_p^*$ is the complex conjugate of $\zeta_p$. The
variables $\zeta_p$ and $\zeta_p^*$ are independent in the same sense
in which an ordinary complex variable and its conjugate are
independent.  The property~\eqref{eq2.1} also holds for the complex
conjugates as well as for mixtures, $\zeta_p \zeta_q^* = - \zeta_q^*
\zeta_p$. There are two different but equivalent ways to interpret
$(\zeta_p^*)^*$. The usual choice in physics is
\begin{equation}
(\zeta_p^*)^* = \zeta_p^{**} = - \zeta_p \ , \qquad p=1,\ldots,k \ ,
\label{eq2.2}
\end{equation}
which has to be supplemented by the rule
\begin{equation}
(\zeta_p\zeta_q \cdots \zeta_r)^* = \zeta_p^*\zeta_q^* \cdots \zeta_r^* \ .
\label{eq2.3}
\end{equation}
There is a concept of reality, since we have
\begin{equation}
(\zeta_p^*\zeta_p)^* = \zeta_p^{**} \zeta_p^* = - \zeta_p \zeta_p^*
                                      = \zeta_p^*\zeta_p \ .
\label{eq2.4}
\end{equation}
Hence, we may interpret $\zeta_p^*\zeta_p$ as the modulus squared of
the complex anticommuting variable $\zeta_p$. Alternatively, one can
use the plus sign in Eq.~\eqref{eq2.2} and reverse the order of the
anticommuting variables on the right hand side of Eq.~\eqref{eq2.3}.
In particular, this also preserves the property~\eqref{eq2.4}.

Because of $\zeta_p^2 = 0$ and since inverse anticommuting
variables do not exist, functions of anticommuting variables can only
be finite polynomials,
\begin{equation}
f(\zeta_1,\ldots,\zeta_k,\zeta_1^*,\ldots,\zeta_k^*) =
\underset{l_p=0,1}{\underset{m_p=0,1}\sum} f_{m_1\cdots m_k l_1\cdots l_k}
  \zeta_1^{m_1}\ldots\zeta_k^{m_k}(\zeta_1^*)^{l_1}\cdots(\zeta_k^*)^{l_k} 
\label{eq2.5}
\end{equation}
with commuting coefficients $f_{m_1\cdots m_k l_1\cdots l_k}$. Thus,
just like functions of matrices, functions of anticommuting variables
are power series. For example, we have
\begin{equation}
\exp\left(a\zeta_p^*\zeta_p\right) = 1 + a\zeta_p^*\zeta_p
                                                             = \frac{1}{1 - a\zeta_p^*\zeta_p} \ .
\label{eq2.6}
\end{equation}
where $a$ is a commuting variable.

\subsection{Vectors and Matrices}
\label{sec2.2}

A supermatrix $\sigma$ is defined via block construction,
\begin{equation}
\sigma = \left[\begin{array}{cc}
                               a     & \mu  \\
                               \nu  & b 
                                \end{array}\right] \ ,
\label{eq2.7}
\end{equation}
where $a$ and $b$ are matrices with ordinary complex commuting entries
while the matrices $\mu$ and $\nu$ have complex anticommuting entries.
Apart from the restriction that the blocks must match, all dimensions
of the matrices are possible. Of particular interest are quadratic
$k_1/k_2 \times k_1/k_2$ supermatrices, that is, $a$ and $b$ have
dimensions $k_1\times k_1$ and $k_2\times k_2$, respectively, $\mu$
and $\nu$ have dimensions $k_1\times k_2$ and $k_2\times k_1$.
A quadratic supermatrix $\sigma$ can have an inverse $\sigma^{-1}$.
Equally important are supervectors, which are defined as special
supermatrices consisting of only one column. As seen in
Eq.~\eqref{eq2.7}, there are two possibilities
\begin{equation}
\psi = \left[\begin{array}{c} z \\
                                                \zeta
                             \end{array}\right] 
\qquad {\rm and} \qquad
\psi = \left[\begin{array}{c} \zeta \\
                                                z
                             \end{array}\right] \ ,
\label{eq2.8}
\end{equation}
where $z$ is a $k_1$ component vector of ordinary complex commuting
entries $z_p$, and $\zeta$ is a $k_2$ component vector of complex
anticommuting entries $\zeta_p$. In the sequel we work with the first
possibility, but everything to be said is valid for the second one
accordingly. The standard rules of matrix addition and multiplication
apply, if everything in Sec.~\ref{sec2.1} is taken into account.
Consider for example the supervector $\psi^\prime$ given by
\begin{equation}
\psi^\prime = \sigma \psi =  \left[\begin{array}{cc}
                                                     a     & \mu  \\
                                                    \nu  & b 
                                                   \end{array}\right]
                                                   \left[\begin{array}{c} z \\
                                                                                         \zeta
                                                                   \end{array}\right] 
                                                 = \left[\begin{array}{c} az+\mu\zeta \\
                                                                                          \nu z+b\zeta 
                                                                               \end{array}\right] \ ,
\label{eq2.9}
\end{equation}
which has the same form as $\psi$. Hence the linear map~\eqref{eq2.9}
transforms commuting into anticommuting degrees of freedom and vice
versa.

The transpose $\sigma^T$ and the Hermitean conjugate $\sigma^\dagger$
are defined as
\begin{equation}
\sigma^T = \left[\begin{array}{cc}
                               a^T     & -\nu^T  \\
                               \mu^T  & b^T 
                                \end{array}\right] 
\qquad {\rm and} \qquad
\sigma^\dagger = (\sigma^T)^* \ .
\label{eq2.10}
\end{equation}
The minus sign in front of $\nu^T$ ensures that
$(\sigma_1\sigma_2)^T=\sigma_2^T\sigma_1^T$ carries over to
supermatrices $\sigma_1$ and $\sigma_2$. Importantly,
$(\sigma^\dagger)^\dagger=\sigma$ always holds, but $(\sigma^T)^T$ is
in general not equal to $\sigma$. As a special application, we define
the scalar product $\psi^\dagger\chi$ where each of the supervectors
$\psi$ and $\chi$ has either the first or the second of the
forms~\eqref{eq2.8}. Because of the reality property~\eqref{eq2.2},
the scalar product $\psi^\dagger\psi$ is real and can be viewed as the
length squared of the supervector $\psi$.

To have cyclic invariance, the supertrace is defined as
\begin{equation}
{\rm str\,}\sigma = {\rm tr\,}a - {\rm tr\,}b 
\label{eq2.10a}
\end{equation}
such that ${\rm str\,}\sigma_1\sigma_2 = {\rm str\,}\sigma_2\sigma_1$
for two different supermatrices $\sigma_1$ and $\sigma_2$.
Correspondingly, the superdeterminant is multiplicative owing to the
definition
\begin{equation}
{\rm sdet\,}\sigma = \frac{\det\left(a-\mu b^{-1} \nu\right)}{\det b}
                                 = \frac{\det a}{\det\left(b-\nu a^{-1} \mu\right)}
\label{eq2.10b}
\end{equation}
for $\det b \neq 0$ such that ${\rm sdet\,}\sigma_1\sigma_2 = {\rm sdet\,}\sigma_1 {\rm
  sdet\,}\sigma_2$.

\subsection{Groups and Symmetric Spaces}
\label{sec2.3}

For and introduction to this topic see chapter 3. Here we only present
the salient features in the context of the supersymmetry method.  The
theory of Lie superalgebras was pioneered by Kac~\cite{Kac77}.
Although the notion of supergroups, particularly Lie supergroups,
seems to be debated in mathematics, a consistent definition from a
physics viewpoint is possible and --- as will become clear later on
--- urgently called for. All supermatrices $u$ which leave the length
of the supervector $\psi$ invariant form the unitary supergroup ${\rm
  U}(k_1/k_2)$.  With $\psi^\prime = u\psi$ we require
$(\psi^\prime)^\dagger \psi^\prime=\psi^\dagger u^\dagger
u\psi=\psi^\dagger \psi$ and the corresponding equation for
$\psi^\prime = u^\dagger\psi$. Hence we conclude
\begin{equation}
u^\dagger u = 1 \ , \quad uu^\dagger = 1 
\qquad {\rm and \ thus} \qquad
u^\dagger = u^{-1} \ .
\label{eq2.12}
\end{equation}
The direct product ${\rm U}(k_1)\times {\rm U}(k_2)$ of ordinary
unitary groups is a trivial subgroup of ${\rm U}(k_1/k_2)$, found by
simply putting all anticommuting variables in $u$ to zero.
Non--trivial subgroups of the unitary supergroup exist as well.
Consider commuting variables, real $w_p , \ p=1,\ldots,k_1$ and
complex $z_{pj}, \ p=1,\ldots,k_1 , \ j=1,2$. We introduce the real
and quaternion--real supervectors
\begin{equation}
\psi = \left[\begin{array}{c} w_1 \\
                                                \vdots \\
                                                w_{k_1} \\
                                                \zeta_1 \\
                                                \zeta_1^*\\
                                                \vdots \\
                                                \zeta_{k_2} \\
                                                \zeta_{k_2}^*
                             \end{array}\right] 
\qquad {\rm and} \qquad
\psi = \left[\begin{array}{cc} z_{11} & -z_{12}^* \\
                                                  z_{12} &   z_{11}^* \\
                                                  \vdots & \vdots \\
                                                  z_{k_11} & -z_{k_12}^* \\
                                                  z_{k_12} &   z_{k_11}^* \\
                                                  \zeta_1^* & -\zeta_1 \\
                                                  \vdots & \vdots \\
                                                  \zeta_{k_2}^* & -\zeta_{k_2}
                             \end{array}\right] \ .
\label{eq2.13}
\end{equation}
The unitary--ortho--symplectic subgroup of the unitary supergroup
leaves the lengths of $\psi$ invariant: ${\rm UOSp}(k_1/2k_2)$ the
length of the first and ${\rm UOSp}(2k_1/k_2)$ the length of the
second supervector in Eq.~\eqref{eq2.13}.  Due to the quaternion
structure in the commuting entries of the second supervector, the
proper scalar product reads ${\rm tr\,}\psi^\dagger\psi$. The trivial
ordinary subgroups are ${\rm O}(k_1)\times{\rm USp}(2k_2)$ $\subset
{\rm UOSp}(k_1/2k_2)$ and ${\rm USp}(2k_1)\times{\rm O}(k_2) \subset
{\rm UOSp}(2k_1/k_2)$, respectively.

As in the ordinary case, non--compact supergroups result from the
requirement that the bilinear form $\psi^\dagger L\psi$ remains
invariant. The metric $L$ is without loss of generality diagonal and
only contains $\pm 1$.  We then have $u^\dagger Lu=L$.

A Hermitean supermatrix $\sigma$ is diagonalized by a supermatrix $u
\in {\rm U}(k_1/k_2)$,
\begin{equation}
  \sigma = u^{-1} s u 
  \qquad {\rm with} \qquad
  s = {\rm diag\,}(s_{11},\ldots,s_{k_11},s_{12},\ldots,s_{k_22}) \ .
\label{eq2.14}
\end{equation}
All eigenvalues $s_{pj}$ are real commuting.  Zirnbauer~\cite{Zir96a}
gave a classification of the Riemannian symmetric superspaces. The
Hermitean symmetric superspace is denoted ${\rm A}|{\rm A}$. Of
interest are also the symmetric superspaces ${\rm AI}|{\rm AII}$ and
${\rm AII}|{\rm AI}$.  The former consists of the $k_1/2k_2 \times
k_1/2k_2$ supermatrices $\sigma=u^{-1} s u$ with $u\in{\rm
  UOSp}(k_1/2k_2)$ and with $s = {\rm
  diag\,}(s_{11},\ldots,s_{k_11},s_{12},s_{12},\ldots,s_{k_22},s_{k_22})$,
the latter of the $2k_1/k_2 \times 2k_1/k_2$ supermatrices $\sigma$
with $u\in{\rm UOSp}(2k_1/k_2)$ and with $s = {\rm
  diag\,}(s_{11},s_{11},\ldots,s_{k_11},s_{k_11},s_{12},\ldots,s_{k_22})$.

\subsection{Derivatives and Integrals}
\label{sec2.4}

Since anticommuting variables cannot be represented by numbers, there
is nothing like a Riemannian integral over anticommuting variables
either. The Berezin integral~\cite{Ber87} is formally defined by
\begin{equation}
\int d\zeta_p = 0
  \qquad {\rm and} \qquad
\int \zeta_p d\zeta_p = \frac{1}{\sqrt{2\pi}} \ ,
\label{eq2.15}
\end{equation}
and accordingly for the complex conjugates $\zeta_p^*$.  The
normalization involving $\sqrt{2\pi}$ is a common, but not the only
convention used.  The differentials $d\zeta_p$ have all the properties
of anticommuting variables collected in Sec.~\ref{sec2.1}. Thus, the
Berezin integral of the function~\eqref{eq2.5} is essentially the
highest order coefficient, more precisely $f_{1\cdots 11\cdots
  1}/(2\pi)^k$ apart from an overall sign determined by the chosen
order of integration.  For example, we have
\begin{equation}
\iint \exp\left(a\zeta_p^*\zeta_p\right) d\zeta_p d\zeta_p^* = \frac{a}{2\pi} \ .
\label{eq2.16}
\end{equation}
This innocent--looking formula is at the heart of the Supersymmetry
method: Anticipating the later discussion, we notice that we would
have found the inverse of the right hand side for commuting
integration variables $z_p$ instead of $\zeta_p$.

One can also define a derivative as the discrete operation
$\partial\zeta_p/\partial\zeta_q=\delta_{pq}$. To avoid ambiguities
with signs, one should distinguish left and right derivatives.
Obviously, derivative and integral coincide apart from factors.
Mathematicians often prefer to think of the Berezin integral as a
derivation. In physics, however, the interpretation as integral is
highly useful as seen when changing variables.  We first consider the
$k_2$ vectors $\zeta$ and $\eta=a\zeta$ of anticommuting variables
where $a$ is an ordinary complex $k_2\times k_2$ matrix. From the
definition~\eqref{eq2.15} we conclude $d[\eta]=\det^{-1}a\, d[\zeta]$.
This makes it plausible that the change of variables in ordinary space
generalizes for an arbitrary transformation $\chi=\chi(\psi)$ of
supervectors in the following manner: Let $y$ be the vector of
commuting and $\eta$ be the vector of anticommuting variables in
$\chi$, then we have
\begin{equation}
d[\chi] = {\rm sdet\,}\frac{\partial\chi}{\partial\psi^T} \, d[\psi]
            = {\rm sdet\,} \left[\begin{array}{cc}
                               \partial y/\partial z^T  & \partial y/\partial\zeta^T \\
                               \partial\eta/\partial z^T & \partial\eta/\partial\zeta^T
                                \end{array}\right]  d[\psi]
\label{eq2.18}
\end{equation}
with $d[\chi]=d[y] d[\eta]$ and $d[\psi]=d[z] d[\zeta]$. The Jacobian
in superspace is referred to as Berezinian.  Absolute value signs are
not needed if we agree to only transform right--handed into
right--handed coordinate systems. Changes of variables in superspace
can lead to boundary contributions which have no analog in ordinary
analysis.  In physics, they are referred to as Efetov--Wegner terms,
see Ref.~\cite{Rot87} for a mathematical discussion.

Importantly, the concept of the $\delta$ function has a meaningful
generalization in superspace. An anticommuting variable $\zeta_p$ acts
formally as $\delta$ function when integrating it with any function
$f(\zeta_p)$, hence $\delta(\zeta_p)=\sqrt{2\pi}\zeta_p$.  More
complicated are expressions of the form $\delta(y-\zeta^\dagger\zeta)$
with an ordinary commuting variable $y$ and a $k$ component vector of
complex anticommuting variables $\zeta$.  To make sense out of it, it
has to be interpreted as
\begin{equation}
\delta(y-\zeta^\dagger\zeta) 
        = \sum_{\kappa=0}^k \frac{(-1)^\kappa}{\kappa!} \delta^{(\kappa)}(y) 
                                                     (\zeta^\dagger\zeta)^\kappa \ .
\label{eq2.19}
\end{equation}
This is a terminating power series, because $(\zeta^\dagger\zeta)^\kappa=0$ for
$\kappa > k$.

\sect{Supersymmetric Representation}
\label{sec3}

Several problems in particle physics would be solved if each Boson had
a Fermionic and each Fermion had a Bosonic partner. A review of this
Supersymmetry can be found in Ref.~\cite{Mar05}. Although
mathematically the same, Supersymmetry in condensed matter physics and
Random Matrix Theory has a completely different interpretation: the
commuting and anticommuting variables do not represent Bosons or
Fermions, that is, physical particles.  Rather, they are highly
convenient bookkeeping devices making it possible to drastically
reduce the number of degrees of freedom in the statistical model.
Since as many commuting as anticommuting variables are involved, one
refers to it as Super--``symmetry'' --- purely formally just like in
particle physics. In 1979, Parisi and Sourlas~\cite{Par79} introduced
superspace concepts to condensed matter physics. Three years later,
Efetov~\cite{Efe82} constructed the supersymmetric non--linear
$\sigma$ model for the field theory describing electron transport in
disordered systems. Efetov and his coworkers developed many of the
tools and contributed a large body of work on
Supersymmetry~\cite{Efe83}.  The first applications of Supersymmetry
to random matrices, that is, in the language of condensed matter
physics, to the zero--dimensional limit of a field theory, were given
by Verbaarschot and Zirnbauer~\cite{Ver85a} and by Verbaarschot,
Zirnbauer and Weidenm\"uller~\cite{Ver85b}. Reviews can be found in
Refs.~\cite{Efe97,Guh98,Mir00}, see also the chapters on chiral Random
Matrix Theory and on scattering.

\subsection{Ensemble Average}
\label{sec3.1}

Using Supersymmetry, the ensemble average in the generating
function~\eqref{eq1.12} is straightforward. We begin with the unitary
case $\beta=2$ and express the determinants as Gaussian integrals
\begin{eqnarray}
\frac{(2\pi)^N}{\det(H - x_p^\pm + J_p)} &=&
     \int d[z_p] \exp\left(iL_pz_p^\dagger(H - x_p^\pm+ J_p)z_p\right)
                                                          \nonumber\\
\frac{\det(H - x_p^\pm + J_p)}{(2\pi)^N} &=&  
     \int d[\zeta_p] \exp\left(i\zeta_p^\dagger(H - x_p^\pm- J_p)\zeta_p\right) 
\label{eq3.1}
\end{eqnarray}
over altogether $k$ vectors $z_p, \ p=1,\ldots,k$ with $N$ complex
commuting entries and $k$ vectors $\zeta_p, \ p=1,\ldots,k$ with $N$
complex anticommuting entries. When integrating over the commuting
variables, the imaginary increment is needed for convergence, for the
integrals over anticommuting variables, convergence is never a
problem. Hence we may write the metric tensor in the form $L={\rm
  diag\,}(L_1,\ldots,L_k,1,\ldots,1)$. Collecting all $H$ dependences,
the ensemble average in Eq.~\eqref{eq1.12} amounts to calculating
\begin{equation}
\Phi(K) = \int d[H]  P(H) \exp\left(i {\rm tr\,} H K\right) \ .
\label{eq3.2}
\end{equation}
where the $N\times N$ matrix $K$ assembles dyadic products of the
vectors $z_p$ and $\zeta_p$,
\begin{equation}
K = \sum_{p=1}^k \left(L_p z_pz_p^\dagger - \zeta_p\zeta_p^\dagger\right) \ .
\label{eq3.3}
\end{equation}
For all $L$, this is a Hermitean matrix $K^\dagger=K$.  

We now turn to the orthogonal case $\beta=1$. At first sight it seems
irrelevant whether $H$ is Hermitean or real--symmetric in the previous
steps. However, the Fourier transform~\eqref{eq3.2} only affects the
real part of $K$, because the imaginary part of $K$ drops out in 
${\rm tr\,} H K$ if $H$ is real--symmetric. Thus, instead of the Gaussian
integrals~\eqref{eq3.1}, we rather use
\begin{eqnarray}
\frac{\pi^N}{\det(H - x_p^\pm + J_p)} &=&
     \int d[w^{(1)}_p] \exp\left(iL_pw^{(1)T}_p(H - x_p^\pm + J_p)w^{(1)}_p\right)
                                                                                                       \nonumber\\
 & & \quad  \int d[w^{(2)}_p] \exp\left(iL_pw^{(2)T}(H - x_p^\pm + J_p)w^{(2)}_p\right)
                                                                                                       \nonumber\\
\frac{\det(H - x_p^\pm + J_p)}{\pi^N} &=&  
     \int d[\zeta_p] \exp\left(i\zeta_p^\dagger(H - x_p^\pm - J_p)\zeta_p\right) 
                                                                                                        \nonumber\\
 & & \qquad\qquad  \exp\left(-i\zeta_p^T(H - x_p^\pm - J_p)\zeta_p^*\right) \ ,
\label{eq3.4}
\end{eqnarray}
where the to $N$ component vectors $w^{(1)}_p$ and $w^{(2)}_p$ have
real entries.  For each $p$, we can construct a $4N$ component
supervector out of $w^{(1)}_p$, $w^{(2)}_p$, $\zeta_p$ and $\zeta_p^*$
whose structure resembles the one of the first of the
supervectors~\eqref{eq2.13}, but with a different number of
components.  Reordering terms, we arrive at the Fourier
transform~\eqref{eq3.2}, but now for real--symmetric $H$ and with 
\begin{equation}
K = \sum_{p=1}^k \left(L_pw^{(1)}_pw^{(1)T}_p + L_pw^{(2)}_pw^{(2)T}_p
                                    - \zeta_p\zeta_p^\dagger  + \zeta_p^*\zeta_p^T\right) \ ,
\label{eq3.5}
\end{equation}
which is $N\times N$ real--symmetric as well. For $\beta=4$, one has
to reformulate the steps in such a way that the corresponding $K$
becomes self--dual Hermitean.

\subsection{Hubbard--Stratonovich Transformation}
\label{sec3.2}

Due to universality, it suffices to assume a Gaussian probability
density function $P(H)\sim\exp(-\beta {\rm tr\,} H^2/2)$ in almost all
applications in condensed matter and many--body physics as well as in
quantum chaos. Hence the random matrices are drawn from the Gaussian
orthogonal (GOE), unitary (GUE) or symplectic ensemble (GSE). The
Fourier transform~\eqref{eq3.2} is then elementary and yields a
Gaussian. The crucial property
\begin{equation}
\Phi(K) = \exp\left( - \frac{1}{2\beta} {\rm tr\,} K^2\right)  
              = \exp\left( - \frac{1}{2\beta} {\rm str\,} B^2\right)
\label{eq3.6}
\end{equation}
holds, where $B$ is supermatrix containing all scalar products of the
vectors to be integrated over. The second equality sign has a purely
algebraic origin. For $\beta=2$, $B$ has dimension $k/k \times k/k$
and reads
\begin{equation}
B = L^{1/2} \left[\begin{array}{cccccc}
                               z_1^\dagger z_1& \cdots & z_1^\dagger z_k &
                               z_1^\dagger \zeta_1& \cdots & z_1^\dagger \zeta_k \\
                               \vdots & & \vdots & \vdots & & \vdots \\
                               z_k^\dagger z_1& \cdots & z_k^\dagger z_k &
                               z_k^\dagger \zeta_1& \cdots & z_k^\dagger \zeta_k \\
                               -\zeta_1^\dagger z_1& \cdots & -\zeta_1^\dagger z_k &
                               -\zeta_1^\dagger \zeta_1& \cdots & -\zeta_1^\dagger \zeta_k  \\
                               \vdots & & \vdots & \vdots & & \vdots \\
                               -\zeta_k^\dagger z_1& \cdots & -\zeta_k^\dagger z_k &
                               -\zeta_k^\dagger \zeta_1& \cdots & -\zeta_k^\dagger \zeta_k
                                \end{array}\right] L^{1/2} \ .
\label{eq3.7}
\end{equation}
While $K$ is Hermitean, the square roots $L^{1/2}$ destroy this
property for $B$, since $L^{1/2}$ can have imaginary units $i$ as
entries, $B$ is Hermitean only for $L=1$. In general, $B$ is in a
deformed (non--compact) form of the symmetric superspace ${\rm A}|{\rm
  A}$. For $\beta=1,4$, the supermatrix $B$ has dimension $2k/2k
\times 2k/2k$ and it is in deformed (non--compact) forms of the
symmetric superspaces ${\rm AI}|{\rm AII}$ and ${\rm AII}|{\rm AI}$,
respectively. We give the explicit forms later on.  The
identity~\eqref{eq3.6} states the keystone of the Supersymmetry
method. The original model in the space of ordinary $N\times N$
matrices is mapped onto a model in space of supermatrices whose
dimension is proportional to $k$, which is the number of arguments in
the $k$--point correlation function.

The Gaussians~\eqref{eq3.6} contain the vectors, that is, their building
blocks to fourth order. To make analytical progress, a Hubbard--Stratonovich
transformation in superspace is used,
\begin{equation}
\exp\left( -  \frac{1}{2\beta} {\rm str\,} B^2\right) = 
       c^{(\beta)}  \int \exp\left( - \frac{\beta}{2} {\rm str\,}(L\sigma)^2\right)
                       \exp\left(i{\rm str\,}L^{1/2}\sigma L^{1/2}B\right) d[\sigma] \ , 
\label{eq3.8}
\end{equation}
where $c^{(2)}=2^{k(k-1)}$ and $c^{(\beta)}=2^{k(4k-3)/2}$ for
$\beta=1,4$. We notice the appearance of the matrices $L$ and
$L^{1/2}$ in~\eqref{eq3.8}. For $L=1$, the supermatrices $\sigma$ and
$B$ have the same symmetries. However, as already observed in the
early eighties for models in ordinary space~\cite{Sch80,Pru82}, this
choice is impossible for $L\neq 1$, because it would render the
integrals divergent. There are two ways out of this problem. One
either constructs a proper explicit parameterization of $\sigma$ or
one inserts the matrices $L$ and $L^{1/2}$ according to~\eqref{eq3.8}.
A mathematically satisfactory understanding of these issues was put
forward only recently in Ref.~\cite{Fyo08}.

Another important remark is called for.  Because of the minus sign in
the supertrace~\eqref{eq2.10a}, a Wick rotation is needed to make the
integral convergent.  It formally amounts to replacing the lower right
block of $\sigma$, that is, $b$ in Eq.~\eqref{eq2.7}, with $ib$.
Apart from that, the metric $L$ is also needed for convergence reason.
Now the vectors appear in second order. They can be ordered in one
large supervector $\Psi$.  For $\beta=2$ it has the form~\eqref{eq2.8}
with $k_1=k_2=kN$, for $\beta=1$ it has the first of the
forms~\eqref{eq2.13} with $k_1=2kN$, $k_2=kN$ and for $\beta=4$ it has
the second of the forms~\eqref{eq2.13} with $k_1=kN$, $k_2=2kN$. The
integral to be done is then seen to be the Gaussian integral in
superspace
\begin{eqnarray}
& & \int \exp\left(i\Psi^\dagger
        \left(L^{1/2}(L^{1/2}\sigma L^{1/2}-x^\pm-J)L^{1/2}\otimes 1_N \right)\Psi\right)d[\Psi]
                                                                                        \nonumber\\
& & \qquad\qquad\qquad\qquad\qquad\qquad 
                                = {\rm sdet\,}^{-N\beta/2\gamma}(\sigma L-x^\pm-J) \ ,
\label{eq3.9}
\end{eqnarray}
where the power $N$ is due to the direct product structure. We
eventually find
\begin{equation}
Z_k(x+J) =  c^{(\beta)}\int \exp\left( - \frac{\beta}{2} {\rm str\,}(L\sigma)^2\right)
                           {\rm sdet\,}^{-N\beta/2\gamma}(\sigma L-x^\pm-J) d[\sigma]
\label{eq3.10}
\end{equation}
as supersymmetric representation of the generating function. The
average over the $N\times N$ ordinary matrix $H$ has been traded for
an average over the matrix $\sigma$ whose dimension is proportional to
$k$, that is, independent of $N$.

\subsection{Matrix $\delta$ Functions and an Alternative Representation}
\label{sec3.3}

In Refs.~\cite{Leh95,Hac95}, a route alternative to the one outlined
in Sec.~\ref{sec3.2} was taken. These authors used matrix $\delta$
functions in superspace and their Fourier representation to express
functions $f(B)$ of the supermatrix $B$ in the form
\begin{equation}
f(B) = \int f(\rho) \delta(\rho-B) d[\rho] 
         = c^{(\beta)2} \int  d[\rho] f(\rho) \int d[\sigma] \exp\left(-i{\rm str\,}\sigma (\rho-B)\right) \ ,
\label{eq3.11}
\end{equation}
where auxiliary integrals over supermatrices $\rho$ and $\sigma$ are
introduced. For simplicity, we only consider $L=1$ here. The function
$\delta(\rho-B)$ is the product of the $\delta$ functions of all
independent variables.  As discussed in Sec.~\ref{sec2.4}, it is
well-defined.  For all functions $f$, formula~\eqref{eq3.11} renders
the integration over the supervector $\Psi$ Gaussian.  When studying
Gaussian averages of ratios of characteristic polynomials,
Fyodorov~\cite{Fyo02} built upon such insights to construct an
alternative representation for the generating function.  He employs a
standard Hubbard--Stratonovich transformation for the lower right
block of the supermatrix $B$ in Eq.~\eqref{eq3.7} which contains the
scalar products $\zeta_p^\dagger \zeta_q$.  He then inserts a $\delta$
function in the space of ordinary matrices to carry out the integrals
over the vectors $z_p$. Although Supersymmetry is used, the generating
function is finally written as an integral over two ordinary matrices
with commuting entries. In this derivation, the Ingham--Siegel
integral
\begin{equation}
I^{({\rm ord})}(R) = \intop_{S>0} \exp\left(-{\rm tr\,}RS\right) {\rm det\,}^{m}S d[S]
        \sim \frac{1}{{\rm det\,}^{m+N}R}
\label{eq3.12}
\end{equation}
for ordinary Hermitean $N\times N$ matrices $R$ and $S$ appears, where
$m\ge 0$.

\subsection{Generalized Hubbard--Stratonovich Transformation
                       and Superbosonization}
\label{sec3.4}

Is Supersymmetry only applicable to Gaussian probability density
functions $P(H)$ ? --- In Ref.~\cite{Hac95}, Supersymmetry and
asymptotic expansions were used to prove universality for arbitrary
$P(H)$. The concept of superbosonization was put forward in
Ref.~\cite{Efe04} and applied in Ref.~\cite{Bun07} to a generalized
Gaussian model comprising a variety of correlations between the matrix
elements. Extending the concept of superbosonization, a full answer to
the question posed above was given in two different but related
approaches in Refs.~\cite{Guh06,Kie09a} and~\cite{Lit08}: An exact
supersymmetric representation exists for arbitrary, well--behaved
$P(H)$. As the equivalence of the two approaches was proven in
Ref.~\cite{Kie09b}, we follow the line of arguing in
Refs.~\cite{Guh06,Kie09a}.  For $\beta=2$, we define the $N \times 2k$
rectangular supermatrix
\begin{equation}
A = \left[ z_1 \cdots z_k \ \zeta_1 \cdots \zeta_k \right] \ ,
\label{eq3.13}
\end{equation}
where the $z_p, \ p=1,\ldots,k$ and $\zeta_p, \ p=1,\ldots,k$ are $N$
component vectors with complex commuting and anticommuting entries,
respectively. We also define the $N \times 4k$  supermatrix
\begin{equation}
A = \left[ z_1 \ z_1^* \cdots  z_k \ z_k^* \ 
                                        \zeta_1 \ \zeta_1^* \cdots \zeta_k \ \zeta_k^* \right] 
\label{eq3.14}
\end{equation}
for $\beta=1$ and eventually the  $2N \times 4k$  supermatrix
\begin{equation}
A = \left[\begin{array}{cc}
                               z_1  & -z_1^*  \\
                               z_1  & z_1^*
                                \end{array} \cdots
               \begin{array}{cc}
                               z_k  & -z_k^*  \\
                               z_k  & z_k^*
                                \end{array} \
               \begin{array}{cc}
                               \zeta_1  & -\zeta_1^*  \\
                               \zeta_1  & \zeta_1^*
                                \end{array} \cdots
               \begin{array}{cc}
                               \zeta_k  & -\zeta_k^*  \\
                               \zeta_k  & \zeta_k^*
                                \end{array} \right]
\label{eq3.15}
\end{equation}
for $\beta=4$. This enables us to write the ordinary matrix $K$ introduced
in Sec.~\ref{sec3.1} and the supermatrix $B$ introduced
in Sec.~\ref{sec3.2} for all $\beta$ in the form
\begin{eqnarray}
K &=& ALA^\dagger = (AL^{1/2})(L^{1/2}A^\dagger) \nonumber\\
B &=& (L^{1/2}A^\dagger)(AL^{1/2}) = L^{1/2}A^\dagger AL^{1/2} \ .
\label{eq3.16}
\end{eqnarray}
For $\beta=2$, we recover Eq.~\eqref{eq3.7}. This algebraic duality
between ordinary and superspace has far--reaching consequences.  One
realizes~\cite{Guh91,Guh06,Lit08} that the integral~\eqref{eq3.2} is
the Fourier transform in matrix space of every, arbitrary probability
density function $P(H)$ and that $\Phi(K)$ is the corresponding
characteristic function. Since we assume that $P(H)$ is rotation
invariant, the same must hold for $\Phi(K)$. Hence, $\Phi(K)$ only
depends on the invariants ${\rm tr\,}K^m, \ m=1,2,3,\ldots$. Due to
cyclic invariance of the trace, the duality~\eqref{eq3.16} implies for
all $m$ the crucial identity
\begin{equation}
{\rm tr\,}K^m = {\rm str\,}B^m \ , \quad {\rm such \ that} \qquad
\Phi(K) = \Phi(B) \ .
\label{eq3.17}
\end{equation}
Hence, viewed as a function of the matrix invariants, $\Phi$ is a
function in ordinary and in superspace. We now employ
formula~\eqref{eq3.11} for $\Phi(K) = \Phi(B)$, do the Gaussian $\Psi$
integrals as usual find for the generating function
\begin{equation}
Z_k(x+J) =   c^{(\beta)2} \int \exp\left(-i{\rm tr\,}(x+J)L\rho\right) \Phi(\rho) I(\rho) d[\rho] 
\label{eq3.18}
\end{equation}
with $I(\rho)$ being a supersymmetric version of the Ingham--Siegel integral.
The supermatrices $\rho$ and $\sigma$ have the same sizes and symmetries as $B$.
A convolution theorem in superspace yields the second form
\begin{equation}
Z_k(x+J) = \int \Pi(\sigma) 
             {\rm sdet\,}^{-N\beta/2\gamma}\left(\sigma L-x^\pm-J\right) d[\sigma] \ ,
\label{eq3.20}
\end{equation}
where $\Pi(\sigma)$ is the superspace Fourier backtransform of the
characteristic function $\Phi(\rho)$.  It plays the r\^ole of the
probability density function for the supersymmetric representation. To
apply these general results for exact calculations, explicit knowledge
of either $\Phi(\rho)$ or $\Pi(\sigma)$ is necessary.

\subsection{More Complicated Models}
\label{sec3.5}

Most advantageously, Supersymmetry allows one to make progress in
important and technically challenging problems beyond the invariant
and factorizing ensembles, for example:
\begin{itemize}
\item Invariant, but non--factorizing ensembles. The probability
  density function $P(H)$ has the property~\eqref{eq1.2}, but not the
  property~\eqref{eq1.7}. They can, in principle, be treated with the
  results of Sec.~\ref{sec3.4}.
\item Sparse or banded random matrices~\cite{Fyo91,Mir91}, see chapter
  23. The probability density function $P(H)$ lacks the invariance
  property~\eqref{eq1.2}.
\item Crossover transitions or external field models.  One is
  interested in the eigenvalue correlations of the matrix
  $H(\alpha)=H^{(0)} + \alpha H$, where $H$ is a random matrix as
  before and where $H^{(0)}$ is either a random matrix with symmetries
  different from $H$ or a fixed matrix. The parameter $\alpha$
  measures the relative strength. As the resolvent in question is now
  $(x_p^--H(\alpha))^{-1}=(x_p^--(H^{(0)} + \alpha H))^{-1}$, we have
  to replace $H$ by $H(\alpha)$ in the determinants in
  Eq.~\eqref{eq1.12}, but not in the probability density function
  $P(H)$ which usually is chosen invariant, see Ref.~\cite{Guh96b}.
\item Scattering theory and other problems, where matrix elements of
  the resolvents enter~\cite{Ver85b}, see chapter 2 on history and
  chapter 34 on scattering. In the Heidelberg formalism~\cite{Mah69},
  scattering is modeled by coupling an effective Hamiltonian which
  describes the interaction zone to the scattering channels. The
  resolvent is then $(x_p^--H+iW)^{-1}$ where the $N\times N$ matrix
  $W$ contains information about the channels.  One has to calculate
  averages of products of matrix elements $[(x_p^--H+iW)^{-1}]_{nm}$.
  To make that feasible, the source variables $J_p$ have to be
  replaced by $N \times N$ source matrices $\tilde{J}_p$ and instead
  of the derivatives~\eqref{eq1.11}, one must calculate derivatives
  with respect to the matrix elements $\tilde{J}_{p,nm}$. The
  probability density function $P(H)$ is unchanged.
\item Field theories for disordered systems, see Refs.~\cite{Efe83,Efe97}.

\end{itemize}
Of course, these and other non--invariant problems can not only be
studied with the Supersymmetry method, other techniques ranging from
perturbative expansions, asymptotic analysis to orthogonal polynomials
supplemented with group integrals are applied as well, see chapters
4., 5. and 6.  Nevertheless, the drastic reduction in the numbers of
degrees of freedom, which is the key feature of Supersymmetry, often
yields precious structural insights into the problem.

\sect{Evaluation and Structural Insights}
\label{sec4}

To evaluate the supersymmetric representation, a large $N$ expansion,
the celebrated non--linear $\sigma$ model, is used in the vast
majority of applications.  We also sketch a method of exact evaluation
which amounts to a diffusion process in superspace.  Throughout, we
focus on the structural aspects.  A survey of the numerous results for
specific systems is beyond the scope of this contribution, we refer
the reader to the reviews in Refs.~\cite{Efe97,Guh98,Mir00}.

\subsection{Non--linear $\sigma$ Model}
\label{sec4.1}

The reduction in the numbers of degrees of freedom is borne out in the
fact that the dimension $N$ of the original random matrix $H$ is an
explicit parameter in Eqs.~\eqref{eq3.10} and~\eqref{eq3.20}. Hence we
can obtain an asymptotic expansion in $1/N$ by means of a saddle point
approximation~\cite{Efe83,Efe97,Ver85a,Ver85b}. This suffices because
one usually is interested in the correlations on the local scale of
the mean level spacing.  Hence, the saddle point approximation goes
hand in hand with the unfolding. The result of this procedure is the
supersymmetric non--linear $\sigma$ model. We consider the two--point
function $k=2$. The integrand in Eq.~\eqref{eq3.10} is written as
$\exp(-F(x+J))$ with the free energy
\begin{equation}
F(x+J) = \frac{\beta}{2} {\rm str\,}(L\sigma)^2
                           + \frac{N\beta}{2\gamma}{\rm str\,}\ln(\sigma L-x^\pm-J) \ ,
\label{eq4.1}
\end{equation}
which is also referred to as Lagrangean.  We introduce center
$\bar{x}=(x_1+x_2)/2$ and difference $\Delta x=x_2-x_1$ of the
arguments. In the large $N$ limit, $\xi=\Delta x/D$ has to be held
fixed where $D\sim 1/\sqrt{N}$ is the local mean level spacing. Hence,
when determining the saddle points, we may set $\Delta x=0$ such that
$x=\bar{x} 1$. Moreover, as we may choose the source variables
arbitrarily small, we set $J=0$ as well. Since all symmetry breaking
terms are gone, the free energy $F(x)$ with $x=\bar{x} 1$ is invariant
under rotations of $\sigma$ which obey the metric $L$. Thus, variation
of $F(x)$ with respect to $\sigma$ yields the scalar equation
\begin{equation}
s_0(\bar{x}-s_0) = \frac{N}{2\gamma} \ , 
\quad {\rm such \ that} \qquad
s_0 = \frac{1}{2}\left(\bar{x} \pm i \sqrt{\frac{2N}{\gamma}-\bar{x}^2}\right) 
\label{eq4.2}
\end{equation}
inside the spectrum, $|\bar{x}|\le\sqrt{2N/\gamma}$.  This is the
famous Pastur equation and its solution $s_0$~\cite{Pas72}. The latter
is proportional to the large $N$ one--point function whose imaginary
part is the Wigner semicircle.  We arrive at the important insight
that the one--point function provides the stable points of the
supersymmetric representation, the correlations on the local scale are
the fluctuations around it. To make this more precise, we recall the
result of Sch\"afer and Wegner~\cite{Sch80} for the non--linear
$\sigma$ model in ordinary space: when doing the large $N$ limit as
sketched above, the imaginary increments of the arguments $x_1$ and
$x_2$ must lie on different sides of the real axis. Otherwise, the
connected part of the two--point function cannot be obtained as seen
from a contour--integral argument.  Hence, the metric $L$ must not be
proportional to the unit matrix, the groups involved are non--compact
and a hyperbolic symmetry is present. This carries over to the
commuting degrees of freedom in superspace~\cite{Efe82}, the groups
are ${\rm U}(1,1/2)$ for $\beta=2$ and ${\rm UOSp}(2,2/4)$ for
$\beta=1,4$. The full saddle point manifold is found to be given by all
non--compact rotations of
$\sigma_0=\bar{x}/2+i\sqrt{2N/\gamma-\bar{x}^2}L/2$ which leave
$F(\bar{x} 1)$ invariant. One parameterizes the group as $u=u_0v$ with
$u_0$ in the direct product ${\rm U}(1/1)\times {\rm U}(1/1)$ for
$\beta=1$ and in ${\rm UOSp(2/2)}\times {\rm UOSp}(2/2)$ for
$\beta=1,4$ and with $v$ in the coset
\begin{eqnarray}
\frac{{\rm U}(1,1/2)}{{\rm U}(1/1)\times {\rm U}(1/1)} 
     \qquad &{\rm for}& \quad  \beta=2 \ , \nonumber\\
\frac{{\rm UOSp}(2,2/4)}{{\rm UOSp(2/2)}\times {\rm UOSp}(2/2)}
     \qquad &{\rm for}& \quad  \beta=1,4 \ .
\label{eq4.3}
\end{eqnarray}
As $u_0$ and $L$ commute, the saddle point manifold is
$v^{-1}\sigma_0v$, that is, essentially $Q=v^{-1}Lv$ with the crucial
property $Q^2=1$. To calculate the correlations, we must re--insert
the symmetry breaking terms $\Delta x$ and $J$ into the free energy.
We put $\sigma=v^{-1}\sigma_0v+\delta\sigma$ and expand to second
order in the variables $\delta\sigma$ which are referred to as massive
modes.  They are integrated out in the generating
function~\eqref{eq3.10} as Gaussian integrals. One is left with
integrals over the coset manifold, that is, over the Goldstone modes.
On the unfolded scale, the two--point correlation
functions~\eqref{eq1.6} acquire the form $1-Y_2(\xi)$.  The two--level
cluster functions read
\begin{equation}
Y_2(\xi) = - {\rm Re\,} \int \exp\left(i\xi{\rm str\,}QL\right) \, 
                                 {\rm str\,}M_1QL  \, {\rm str\,}M_2QL \, d\mu(Q) \ ,
\label{eq4.4}
\end{equation}
where $d\mu(Q)$ is the invariant measure on the saddle point manifold.
The matrices $M_i, \ i=1,2$ result from the derivatives with respect
to the source variables, $M_i$ is found by formally setting $J_i=1$
and $J_l=0, \ l\ne i,$ in the matrix $J$. The
expressions~\eqref{eq4.4} can be reduced to two radial integrals on
the coset manifold for the GUE and to three such integrals for GOE and
GSE. Efetov~\cite{Efe83} discovered Eq.~\eqref{eq4.4} when taking the
zero--dimensional limit of his supersymmetric non--linear $\sigma$
model for electron transport in disordered mesoscopic systems. He
thereby established a most fruitful link between Random Matrix Theory
and mesoscopic physics. 

The non--linear $\sigma$ model, particularly its mathematical aspects,
was recently reviewed in Ref.~\cite{Zir06}.

\subsection{Eigenvalues and Diffusion in Superspace}
\label{sec4.2}

The supersymmetric representation and Fyodorov's alternative
representation of Sec.~\ref{sec3} are exact for finite $N$. In some
situations, it is indeed possible and advantageous to evaluate them
without using the non--linear $\sigma$ model. It has been shown for
the supersymmetric representation~\cite{Guh06,Kie09a} that the
imaginary increments of the arguments may then all lie on the same
side of the real axis. We have $L=1$ and all groups are compact.  As
we aim at structural aspects, we consider the crossover transitions
involving $H(\alpha)=H^{(0)} + \alpha H$ as discussed in
Sec.~\ref{sec3.5}. We introduce the fictitious time $t=\alpha^2/2$.
Dyson~\cite{Dys62,Dys72} showed that the eigenvalues of $H(t)$ follow
a Brownian motion in $t$ which implies that their probability density
function is propagated by a diffusion equation.  Without any loss of
information, Supersymmetry reduces this stochastic process to a
Brownian motion in a much smaller space~\cite{Guh96b}. It is precisely
the radial part of the Riemannian symmetric superspaces discussed in
Sec.~\ref{sec2.3}. The quantity propagated is then the generating
function $Z_k(x+J,t)$ of the correlations. The initial condition
\begin{equation}
Z_k^{(0)}(s) =  \int P^{(0)}(H^{(0)})
                           {\rm sdet\,}^{-1}(s\otimes 1 + 1 \otimes H^{(0)}) d[H^{(0)}] 
\label{eq4.5}
\end{equation}
is arbitrary, it includes ensembles, but also a fixed matrix $H^{(0)}$
if the probability density $P^{(0)}(H^{(0)})$ is chosen accordingly.
Due to the direct product structure, $Z_k^{(0)}(s)$ is rotation
invariant.  The diagonal matrix $s$ is in the above mentioned radial
space, such that $\sigma=u^{-1}su$, see Sec.~\ref{sec2.3}. For
$\beta=2$, this space coincides with $x+J$, for $\beta=1,4$, it is
slightly larger.  The generating function is then a convolution in the
radial space,
\begin{equation}
Z_k(r,t) = \int \Gamma_k(s,r,t) Z_k^{(0)}(s) B_k(s) d[s] \ .
\label{eq4.6}
\end{equation}
When going to eigenvalue--angle coordinates $\sigma=u^{-1}su$
Berezinians $B_k(s)$ occur analogous to $|\Delta_N(X)|^\beta$ in
ordinary space. The propagator is the supergroup integral
\begin{equation}
\Gamma_k(s,r,t) = c^{(\beta)} \exp\left( - \frac{\beta}{4t} {\rm str\,}(s^2+r^2)\right) 
                    \int \exp\left( - \frac{\beta}{2t} {\rm str\,}u^{-1}sur\right) d\mu(u) \ .
\label{eq4.7}
\end{equation}
For $\beta=2$ and all $k$, this integral is known
explicitly~\cite{Guh91,Guh96a}. Unfortunately, for $\beta=1,4$ the
available result~\cite{Guh02} is handy only for $k=1$ but cumbersome
for $k=2$.

It is a remarkable inherent feature of Supersymmetry that the
propagator and thus the diffusion process of $Z_k(r,t)$ on the
original scale in $t$ carries over unchanged to the unfolded scale
when introducing the proper time $\tau=t/D^2$. The initial condition
is the unfolded large $N$ limit of $Z_k^{(0)}(s)$.  Moreover and in
contrast to the hierarchical equations for the correlation
functions~\cite{Fre88}, the Brownian motion in superspace for the
generating functions is diagonal in $k$.

\sect{Circular Ensembles and Color--Flavor Trans\-for\-ma\-tion}
\label{sec5}

In many physics applications, the random matrices $H$ model
Hamiltonians, implying that they are either real symmetric, Hermitean
or quaternion real.  Mathematically speaking, they are in the
non--compact forms of the corresponding symmetric spaces.  However, if
one aims at modeling scattering, it is often useful to work with
random unitary matrices $S$, taken from the compact forms of these
symmetric spaces~\cite{Dys62}. These are ${\rm U}(N)/{\rm O}(N)$,
${\rm U}(N)$ and ${\rm U}(2N)/{\rm Sp}(2N)$, respectively, leading to
the circular orthogonal, unitary and symplectic ensemble COE, CUE and
CSE which are labeled $\beta=1,2,4$. The phase angles of $S$ play the
same r\^ole as the eigenvalues of $H$. Due to the compactness, no
Gaussian or other confining function is needed and the probability
density function is just the invariant measure on the symmetric space
in question. On the local scale of the mean level spacing, the
correlations coincide with those of the Gaussian
ensembles~\cite{Dys62,Meh04}.

Zirnbauer~\cite{Zir96b} showed how to apply Supersymmetry to the
circular ensembles. His approach works for all three symmetry classes,
but for simplicity we only discuss the CUE which consists of the
unitary matrices $S=U \in {\rm U}(N)$. Consider the generating
function
\begin{equation}
Z_{k_+k_-}(\vartheta,\varphi) = \int d\mu(U) 
    \prod_{p=1}^{k_+} \frac{\det\left(1-\exp(i\varphi_{+p})U\right)}
                                              {\det\left(1-\exp(i\vartheta_{+p})U\right)}
    \prod_{q=1}^{k_-} \frac{\det\left(1-\exp(i\varphi_{-q})U^\dagger\right)}
                                              {\det\left(1-\exp(i\vartheta_{-q})U^\dagger\right)} \ ,
\label{eq5.1}
\end{equation}
where $d\mu(U)$ is the invariant measure on ${\rm U}(N)$.  To derive
the correlation functions $R_k(\vartheta_1,\ldots,\vartheta_k)$ one
sets $k_+=k_-=k$, takes derivatives with respect to the variables
$\varphi_{\pm p}$ and puts certain combinations of variables
$\varphi_{\pm p}$ and $\vartheta_{\pm p}$ equal. The variables
$\vartheta_{\pm p}$ in Eq.~\eqref{eq5.1} have small imaginary
increments to prevent $Z_{k_+k_-}(\vartheta,\phi)$ from becoming
singular.

Since the Hubbard--Stratonovich transformation of Sec.~\ref{sec3.2}
cannot be employed to construct a supersymmetric representation of
$Z_{k_+k_-}(\vartheta,\phi)$, Zirnbauer~\cite{Zir96b} developed the
color--flavor transformation based on the identity
\begin{eqnarray}
& & \int d\mu(U) \exp\left(\Psi_{+pj}^{n*} U^{nm} \Psi_{+pj}^m + 
                                               \Psi_{+qj}^{n*} U^{nm*} \Psi_{+qj}^m\right) =
                                                    \nonumber\\
& & \int d[\Lambda]d[\tilde{\Lambda}] 
                       {\rm sdet\,}^N\left(1-\tilde{\Lambda}\Lambda\right)
                \exp\left(\Psi_{+pj}^{n*}\Lambda_{pjql}\Psi_{-ql}^n +
                                \Psi_{-ql}^{m*}\tilde{\Lambda}_{qlpj}\Psi_{+pj}^m\right)
\label{eq5.2}
\end{eqnarray}
which transforms an integral over the ordinary group ${\rm U}(N)$ into
an integral over $k_+/k_+ \times k_-/k_-$ rectangular supermatrices
$\Lambda$ and $\tilde{\Lambda}$ parameterizing the coset space ${\rm
  U}(k_++k_-/k_++k_-)/{\rm U}(k_+/k_+)\times {\rm U}(k_-/k_-)$.  The
integrals depend on the supertensor $\Psi$ with components
$\Psi_{+pj}^n$.  The indices $n,m=1,\ldots,N$ label the elements of
$U$ in ordinary space. The indices $pj$ and $ql$ are superspace
indices with $p=1,\ldots,k_+, \ q=1,\ldots,k_-$ and with $j,l=1,2$
labeling the four blocks of the supermatrices, see Eq.~\eqref{eq2.7}.
Summation convention applies.  The superfields $\Psi$ are used to
express the determinants in Eq.~\eqref{eq5.1} as Gaussian integrals.
After the color--flavor transformation, they are integrated out again.

As the name indicates, the color--flavor transformation is naturally
suited for applications in lattice gauge theories where $U$ is in the
color gauge group, see Ref.~\cite{Nag01}. In Ref.~\cite{Wei05}, the
color--flavor transformation was derived for the gauge group ${\rm
  SU}(N)$ relevant in lattice quantum chromodynamics.

\sect{Concluding Remarks}
\label{sec6}

Apart from the wealth of results for specific physics systems which
had to be left out, some important conceptual issues could not be
discussed here either due to lack of space: we only mentioned
applications of the other Riemannian symmetric
superspaces~\cite{Zir96a} to Andreev scattering and chiral Random
Matrix Theory. As random matrix approaches are now ubiquitous in
physics and beyond, one may also expect that the Supersymmetry method
spreads out accordingly.  From a mathematical viewpoint, various
aspects deserve further clarifying studies, in the present context
most noticeably the theory of supergroups and harmonic analysis on
superspaces.

\ \\
{\sc Acknowledgements}: 

I thank Mario Kieburg and Heiner Kohler for helpful discussions. I
acknowledge support from Deutsche Forschungsgemeinschaft within
Sonderforschungsbereich Transregio 12 ``Symmetries and Universality in
Mesoscopic Systems''.

\end{document}